\documentclass{aa}
\usepackage{graphicx}
\newcommand{\dfbump}{$\Delta\mathrm{F555W}^{\mathrm{bump}}_{\mathrm{HB}}$}
\newcommand{\dvbump}{$\Delta{V}^{\mathrm{bump}}_{\mathrm{HB}}$}
\newcommand{\rbump}{$R_{\mathrm{bump}}$}
\newcommand{\feh}{\mathrm{[Fe/H]}}
\begin{document}
\title{The Red Giant Branch Luminosity Function Bump
\thanks{Based on observations with the NASA/ESA {\it Hubble Space
Telescope},
obtained at the Space Telescope Science Institute, which is operated by
AURA,
Inc., under NASA contract NAS5-26555, and on observations retrieved with
the
ESO ST-ECF Archive.}
}

\author{
M. Riello \inst{1,2}
\and
S. Cassisi \inst{3}
\and
G. Piotto \inst{2}
\and
A. Recio-Blanco \inst{2}
\and
F. De Angeli \inst{2}
\and
M. Salaris \inst{4}
\and
A. Pietrinferni \inst{3}
\and
G. Bono \inst{5}
\and
M. Zoccali\inst{6}
}

\offprints{M. Riello \email{riello@pd.astro.it}}

\institute{
INAF - Osservatorio Astronomico di Padova, vicolo dell'Osservatorio 5,
I--35122 Padova, Italy
\and
Dipartimento di Astronomia, Universit\`a di Padova, vicolo
dell'Osservatorio 2, I--35122 Padova, Italy
\and
INAF - Osservatorio Astronomico di Collurania, Via M. Maggini, Teramo,
I--64100, Italy
\and
Astrophysics Research Institute, Liverpool John Moores University, Twelve
Quays House, Birkenhead, CH41 1LD, UK
\and
INAF - Osservatorio Astronomico di Roma, Via Frascati 33, 00040 Monte Porzio
Catone, Italy
\and
European Southern Observatory, Karl Schwarzschild Str. 2, D-85748, Garching
b. M\"unchen, Germany.
}
\date{Received xxx/ Accepted xxx}
\abstract{We present observational estimates of the magnitude difference 
between the luminosity function red giant branch bump and the horizontal 
branch (\dfbump), and of star counts in the bump region (\rbump), for a 
sample of 54 Galactic globular clusters observed by the HST. The large 
sample of stars resolved in each cluster, and the high photometric accuracy 
of the data allowed us to detect the bump also in a number of metal poor 
clusters.

To reduce the photometric uncertainties, empirical values are compared with
theoretical predictions obtained from a set of updated canonical stellar
evolution models which have been transformed directly into the HST flight
system.  We found an overall qualitative agreement between theory and
observations. Quantitative estimates of the confidence level are hampered
by current uncertainties on the globular cluster metallicity scale, and by 
the strong dependence of \dfbump\ on the cluster metallicity.
In case of the \rbump\ parameter, which is only weakly affected by the 
metallicity, we find a very good quantitative agreement between theoretical 
canonical models and observations. For our full cluster sample the average 
difference between predicted and observed \rbump\ values is practically 
negligible, and ranges from $-$0.002 to $-$0.028, depending on the
employed metallicity scale. The observed dispersion around these values is 
entirely consistent with the observational errors on \rbump.
As a comparison, the value of \rbump\ predicted by theory in case of 
spurious bump detections due to Poisson noise in the stellar counts would 
be $\sim$0.10 smaller than the observed ones.

We have also tested the influence on the predicted \dfbump\ and \rbump\ 
values of an He-enriched component in the cluster stellar population, as 
recently suggested by D'Antona et al.~(\cite{d02}). We find that, under 
reasonable assumptions concerning the size of this He-enriched population 
and the degree of enrichment, the predicted \dfbump\ and \rbump\ values are 
only marginally affected.
}
\titlerunning{The Red Giant Branch Bump}
\authorrunning{M. Riello et al.}
\maketitle
%

\section{Introduction}

Stellar evolution models supply fundamental tools which enable one to 
constrain the structure of the Milky Way, the properties of extragalactic 
stellar systems, and the early evolution of the universe.
During the last few years the substantial increase in the spatial resolution 
provided by the Hubble Space Telescope (HST), as well as the advent of wide 
field imagers in ground-based telescopes provided homogeneous and accurate 
photometry for large samples of stars in Galactic Globular Clusters (GGCs). 
This has made possible a thorough comparison between theoretical models of 
low mass stars and observations, over a broad metallicity range.
This kind of comparisons play a crucial role in modern stellar astrophysics,
because they provide stringent tests for the accuracy of current stellar 
evolution models (Renzini \& Fusi Pecci \cite{rfp88}; Castellani 1999).
    
In this context, the Red Giant Branch (RGB) luminosity function (LF) of 
GGCs is an important tool to test the chemical stratification inside the 
stellar envelopes (Renzini \& Fusi Pecci \cite{rfp88}).
The most interesting feature of the RGB LF is the occurrence of a local 
maximum in the luminosity distribution of RGB stars, which appears as a 
bump in the differential LF, and as a change in the slope of the cumulative 
LF.
According to Thomas~(\cite{thomas}) and Iben~(\cite{iben}) this feature is 
caused by the sudden increase of H-abundance left over by the surface 
convection upon reaching its maximum inward extension at the base of the 
RGB. When the advancing H-burning shell encounters this discontinuity, its 
efficiency is affected (sudden increase of the available fuel), causing a 
temporary drop of the surface luminosity. After some time the thermal  
equilibrium is restored and the surface luminosity starts to increase again.
As a consequence, the stars cross the same luminosity interval three times, 
and this occurrence shows up as a characteristic peak in the differential LF 
of RGB stars. Moreover, since the H-profile before and after the 
discontinuity is different, the rate of advance of the H-burning shell 
changes when the discontinuity is crossed, thus causing a change in the 
slope of the cumulative LF.

The brightness of the RGB bump is therefore related to the location of
this H-abundance discontinuity, in the sense that the deeper the chemical
discontinuity is located, the fainter is the bump luminosity.
A comparison between the predicted bump luminosity and the observations
allows a direct check of how well theoretical models for RGB stars predict 
the extension of convective regions in the stellar envelope; following 
Fusi Pecci et al.~(\cite{fp90}), the observed magnitude difference between 
the RGB bump and the HB at the RR Lyrae instability strip (\dvbump) is 
usually employed in order to test the theoretical predictions for the bump 
brightness.  
This quantity presents several advantages from the observational point of 
view (see Fusi Pecci et al.~\cite{fp90}, and Salaris et al.~\cite{scw02}) 
and it is empirically well-defined because it does not depend on a previous 
knowledge of the cluster distance.

Dating back to its first detection in the LF of the GGC 47~Tuc by King,
Da Costa \& Demarque~(\cite{king}), the RGB bump has been the subject
of several theoretical and observational investigations (Fusi Pecci
et al.~\cite{fp90}, Cassisi \& Salaris~\cite{cs97}, Alves \&
Sarajedini~\cite{alves}, Ferraro et al.~\cite{f99}, Zoccali et
al.~\cite{z99}, Bergbusch \& Vandenberg~\cite{bergbusch},
Salaris, Cassisi \& Weiss~\cite{scw02} and references therein).  
However, a sound quantitative comparison between theory and observations 
was hampered by the size of the observed stellar samples along the RGB, 
and by the heterogeneity of the datasets.
This problem was even more severe for the most metal-poor GGCs, since
the RGB evolutionary timescales is significantly shorter in metal-poor
than in metal-rich stars. However, Zoccali et al. (1999, hereinafter Z99),
using a homogeneous set of data collected with HST, firmly detected the
RGB bump in a large sample of GGCs covering a wide metallicity range.

This is the sixth in a series of papers aimed at investigating the RGB 
bump in the LF of GGCs. On the basis of a detailed set of evolutionary 
models Cassisi \& Salaris (\cite{cs97}) showed that the predicted \dvbump
was in agreement with empirical estimates of GGCs with accurate 
spectroscopic measurements of heavy-element abundances. This finding was 
further strengthened by the evidence that the \dvbump\ is only marginally 
affected by atomic diffusion (Cassisi, Degl'Innocenti \& Salaris 
\cite{cds97}). A more detailed comparison was provided by Z99 who found 
that theory and observations do agree at the level of $\approx0.1$ mag.    

A different kind of analysis was performed by Bono et al. (\cite{bono},
hereinafter B01), who compared the theoretical evolutionary lifetimes
during the crossing of the H-discontinuity with the star counts across 
the RGB bump, a quantity that is sensitive to possible extra-mixing 
processes below the formal convective boundary. B01 have shown that star 
counts in the bump region are sensitive to the size of the jump in the 
H-profile left over when the envelope starts to recede, after achieving 
its maximum inward extension.
Deep mixing phenomena (prior to the bump stage) able to dredge up an 
appreciable amount of He would change the size of the H-abundance jump, 
thus modifying the star counts in the bump region.
The \rbump\ parameter was defined as the ratio between the star counts in 
the bump region $\mathrm{V}_{\mathrm{bump}}\pm 0.4$ and the star counts in 
the normalization region
$\mathrm{V}_{\mathrm{bump}}+0.5 < \mathrm{V}_{\mathrm{bump}}
< \mathrm{V}_{\mathrm{bump}}+ 1.5$. 
The reasons for this choice of the bump and the normalization regions are 
that the former should be large enough to include all bump stars, while 
the latter is required to normalize the total number of stars sampled in 
each cluster.
The comparison between theory and observations performed by B01 disclosed
that the occurrence of a substantial efficiency of non-canonical
extra-mixing before the RGB bump could be excluded.
On the basis of theoretical arguments, Cassisi, Salaris \& Bono (\cite{csb02}) 
more recently suggested to use the shape of the LF bump as a complementary
diagnostic of partial mixing processes at the base of the outer convective 
zone, since it is sensitive to the shape of the H-discontinuity.

The key differences between this investigation and Z99 are the following: 
{\em i)} we adopted the sample of GGCs presented by Piotto et al.
(\cite{snapshot}) and for 54 of them we detected the RGB bump. This sample
is approximately 30\% larger than the sample by Z99;
{\em ii)} we adopted the more robust HST flight photometric system
(without reddening corrections) instead of the Johnson de-reddened bands.
This approach avoids deceptive errors in the estimate of the visual
magnitudes of both RGB bump and HB. As a matter of fact, the calibration  
to the standard Johnson system (Dolphin \cite{dolphin}) requires the
knowledge of the reddening. Therefore, the accuracy of the equivalent
Johnson magnitudes depends on the accuracy of the adopted cluster reddening 
(see Piotto et al. \cite{snapshot});
{\em iii)} the uncertainties of current parameters have been estimated
using the completeness curves and the photometric errors obtained from a 
large number of artificial star experiments (see \S \ref{database});
{\em iv)} the comparison between theory and observations is based on an 
updated theoretical framework.

The main aim of this investigation is to provide new homogeneous 
measurements of the RGB bump, and in turn of its brightness with respect to 
the HB, for a sizable sample of GGCs that cover a wide range in metallicity
($-2.2\leq\textrm{[Fe/H]}\leq-0.3$), and to compare these measurements
with predictions based on a new set of evolutionary models.

In the next section we present the observational database, and we briefly 
discuss the procedures adopted for the reduction and the calibration of 
the data. In this section we also outline the approach adopted to estimate 
the \dfbump\ and the $R_{\rm bump}$ parameters.
In \S3, we present the new theoretical models adopted in this investigation, 
while in \S4 we compare theory and observations. Conclusions and future 
developments are discussed in \S5.

\section{The cluster database}
\label{database}

We exploited our large photometric database of 74 GGCs observed in the 
HST \(B\) (F439W) and \(V\) (F555W) bands with the WFPC2 (Piotto et al. 
\cite{snapshot}) to measure the \dfbump\ and \rbump\ parameters, that are 
related to the position and extension of the RGB bump. The observations, 
pre-processing, photometric reduction, and calibration of the instrumental 
magnitudes to the HST flight system as well as the artificial star experiments
performed to derive star count completeness are described in Piotto et al. 
(\cite{snapshot}). All photometric data have been processed following the 
same steps: after the pre-processing, the instrumental photometry for each 
cluster was obtained with DAOPHOT II/ALLFRAME; the correction for the CTE 
effect and the calibration to the flight system was performed following the
prescriptions by Dolphin (\cite{dolphin}).

For each cluster we measured a number of stars that ranges from a few
thousands in the less massive clusters, to $\approx 47,000$ in NGC~6388.
This and the high internal photometric accuracy, (the typical photometric 
error at the bump and ZAHB level ranges from a few $10^{-3}$ to a few 
$10^{-2}$ mag) allowed us to detect the RGB bump even in metal-poor clusters.

\begin{table*}[!t]
\begin{center}
\caption{Data for the 54 GGCs in our sample. The 26 clusters with more 
than 85 stars in the RGB Bump area are marked with an asterisk.}
\label{tabump}
\begin{tabular}{lcccccccc}
\hline
\hline
\noalign{\smallskip}
Object & $[M/H]_{CG97}$ & $[M/H]_{ZW84}$ &
$\mathrm{F555W}_{\mathrm{bump}}$ &
\dfbump\ & $N_B$ & $N_N$ & \rbump \\
\noalign{\smallskip}
(1) & (2) & (3) & (4) & (5) & (6) & (7) & (8) \\
\noalign{\smallskip}
\hline
ic4499                 & -1.06 & -1.29 & 17.42$\;\pm\;$0.04 &
-0.34$\;\pm\;$0.08 &   20 &   62 & 0.323$\;\pm\;$0.083 \\
n0104$^\ast$  (47 Tuc) & -0.56 & -0.57 & 14.58$\;\pm\;$0.03 &
 0.32$\;\pm\;$0.08 &  295 &  480 & 0.615$\;\pm\;$0.046 \\
n0362$^\ast$           & -0.94 & -1.06 & 15.46$\;\pm\;$0.04 &
-0.05$\;\pm\;$0.10 &  129 &  268 & 0.481$\;\pm\;$0.052 \\
n1261                  & -0.89 & -1.10 & 16.76$\;\pm\;$0.04 &
-0.09$\;\pm\;$0.08 &   70 &  172 & 0.407$\;\pm\;$0.058 \\
n1851$^\ast$           & -0.93 & -1.15 & 16.12$\;\pm\;$0.04 &
-0.04$\;\pm\;$0.10 &  147 &  320 & 0.459$\;\pm\;$0.046 \\
n1904  (M~79)          & -1.16 & -1.48 & 16.00$\;\pm\;$0.08 &
-0.26$\;\pm\;$0.11 &   69 &  130 & 0.531$\;\pm\;$0.079 \\
n2808$^\ast$           & -0.94 & -1.16 & 16.29$\;\pm\;$0.03 &
-0.07$\;\pm\;$0.08 &  387 &  818 & 0.473$\;\pm\;$0.029 \\
n4590  (M~68)          & -1.78 & -1.88 & 15.28$\;\pm\;$0.04 &
-0.45$\;\pm\;$0.09 &   29 &   38 & 0.763$\;\pm\;$0.188 \\
n4833                  & -1.37 & -1.65 & 15.21$\;\pm\;$0.03 &
-0.49$\;\pm\;$0.08 &   27 &   67 & 0.403$\;\pm\;$0.092 \\
n5024$^\ast$  (M~53)   & -1.68 & -1.83 & 16.63$\;\pm\;$0.03 &
-0.20$\;\pm\;$0.08 &   89 &  224 & 0.397$\;\pm\;$0.050 \\
n5634                  & -1.40 & -1.61 & 17.43$\;\pm\;$0.03 &
-0.26$\;\pm\;$0.08 &   63 &  114 & 0.553$\;\pm\;$0.087 \\
n5694                  & -1.52 & -1.71 & 18.25$\;\pm\;$0.04 &
-0.28$\;\pm\;$0.08 &   84 &  181 & 0.464$\;\pm\;$0.061 \\
n5824$^\ast$           & -1.46 & -1.66 & 18.15$\;\pm\;$0.04 &
-0.38$\;\pm\;$0.08 &  211 &  431 & 0.490$\;\pm\;$0.041 \\
n5904         (M~5)    & -0.90 & -1.19 & 15.01$\;\pm\;$0.03 &
-0.19$\;\pm\;$0.08 &   84 &  168 & 0.500$\;\pm\;$0.067 \\
n5927$^\ast$           & -0.48 & -0.16 & 17.34$\;\pm\;$0.03 &
 0.45$\;\pm\;$0.08 &  131 &  253 & 0.518$\;\pm\;$0.056 \\
n5946                  & -0.94 & -1.16 & 17.33$\;\pm\;$0.04 &
-0.27$\;\pm\;$0.09 &   58 &  126 & 0.460$\;\pm\;$0.073 \\
n5986$^\ast$           & -1.23 & -1.46 & 16.45$\;\pm\;$0.04 &
-0.26$\;\pm\;$0.08 &  103 &  205 & 0.502$\;\pm\;$0.061 \\
n6093$^\ast$  (M~80)   & -1.24 & -1.47 & 16.03$\;\pm\;$0.03 &
-0.32$\;\pm\;$0.08 &  133 &  280 & 0.474$\;\pm\;$0.050 \\
n6139$^\ast$           & -1.21 & -1.44 & 17.95$\;\pm\;$0.04 &
-0.10$\;\pm\;$0.09 &  141 &  297 & 0.474$\;\pm\;$0.049 \\
n6171  (M~107)         & -0.73 & -0.85 & 15.88$\;\pm\;$0.04 &
 0.09$\;\pm\;$0.08 &   28 &   51 & 0.549$\;\pm\;$0.129 \\
n6205$^\ast$  (M~13)   & -1.18 & -1.44 & 14.75$\;\pm\;$0.04 &
-0.31$\;\pm\;$0.08 &  112 &  298 & 0.377$\;\pm\;$0.042 \\
n6218  (M~12)          & -1.16 & -1.40 & 14.83$\;\pm\;$0.04 &
-0.27$\;\pm\;$0.08 &   27 &   52 & 0.519$\;\pm\;$0.123 \\
n6229$^\ast$           & -1.09 & -1.33 & 17.99$\;\pm\;$0.04 &
-0.11$\;\pm\;$0.09 &  132 &  283 & 0.466$\;\pm\;$0.049 \\
n6235                  & -0.96 & -1.19 & 17.30$\;\pm\;$0.04 &
 0.30$\;\pm\;$0.09 &   21 &   68 & 0.309$\;\pm\;$0.077 \\
n6266$^\ast$  (M~62)   & -0.86 & -1.07 & 16.27$\;\pm\;$0.03 &
-0.03$\;\pm\;$0.08 &  231 &  521 & 0.443$\;\pm\;$0.035 \\
n6284                  & -0.96 & -1.19 & 17.44$\;\pm\;$0.03 &
-0.06$\;\pm\;$0.08 &   83 &  171 & 0.485$\;\pm\;$0.065 \\
n6293                  & -1.52 & -1.71 & 16.05$\;\pm\;$0.04 &
-0.43$\;\pm\;$0.08 &   60 &   98 & 0.612$\;\pm\;$0.100 \\
n6342                  & -0.55 & -0.48 & 17.71$\;\pm\;$0.04 &
 0.54$\;\pm\;$0.09 &   54 &  115 & 0.470$\;\pm\;$0.078 \\
n6355                  & -1.06 & -1.29 & 17.96$\;\pm\;$0.04 &
 0.17$\;\pm\;$0.08 &   46 &  146 & 0.315$\;\pm\;$0.053 \\
n6356$^\ast$           & -0.55 & -0.48 & 18.18$\;\pm\;$0.05 &
 0.43$\;\pm\;$0.09 &  309 &  618 & 0.501$\;\pm\;$0.035 \\
n6362                  & -0.82 & -0.87 & 15.53$\;\pm\;$0.04 &
 0.11$\;\pm\;$0.09 &   29 &   50 & 0.580$\;\pm\;$0.135 \\
n6388$^\ast$           & -0.60 & -0.60 & 17.79$\;\pm\;$0.05 &
 0.40$\;\pm\;$0.10 &  852 & 1918 & 0.444$\;\pm\;$0.018 \\
n6402$^\ast$  (M~14)   & -0.95 & -1.18 & 17.36$\;\pm\;$0.03 &
-0.04$\;\pm\;$0.08 &  169 &  375 & 0.451$\;\pm\;$0.042 \\
n6441$^\ast$           & -0.54 & -0.45 & 18.47$\;\pm\;$0.05 &
 0.41$\;\pm\;$0.09 & 1006 & 2194 & 0.459$\;\pm\;$0.018 \\
n6453$^\ast$           & -1.08 & -1.32 & 17.61$\;\pm\;$0.04 &
-0.20$\;\pm\;$0.12 &   94 &  196 & 0.478$\;\pm\;$0.060 \\
n6522$^\ast$           & -1.00 & -1.23 & 16.85$\;\pm\;$0.04 &
 0.05$\;\pm\;$0.09 &  103 &  169 & 0.609$\;\pm\;$0.076 \\
n6544                  & -1.11 & -1.35 & 15.34$\;\pm\;$0.04 &
 0.08$\;\pm\;$0.08 &   15 &   32 & 0.466$\;\pm\;$0.145 \\
n6569$^\ast$           & -0.66 & -0.72 & 17.87$\;\pm\;$0.04 &
 0.15$\;\pm\;$0.09 &  151 &  316 & 0.478$\;\pm\;$0.047 \\
n6584                  & -1.09 & -1.33 & 16.48$\;\pm\;$0.04 &
-0.11$\;\pm\;$0.09 &   43 &  101 & 0.426$\;\pm\;$0.078 \\
n6624$^\ast$           & -0.49 & -0.21 & 16.74$\;\pm\;$0.05 &
 0.53$\;\pm\;$0.10 &  124 &  190 & 0.653$\;\pm\;$0.075 \\
n6637$^\ast$  (M~69)   & -0.55 & -0.45 & 16.47$\;\pm\;$0.05 &
 0.33$\;\pm\;$0.09 &  104 &  196 & 0.531$\;\pm\;$0.064 \\
n6638$^\ast$           & -0.83 & -0.94 & 17.14$\;\pm\;$0.04 &
 0.17$\;\pm\;$0.10 &   94 &  172 & 0.549$\;\pm\;$0.070 \\
n6642                  & -0.87 & -1.08 & 16.63$\;\pm\;$0.04 &
-0.07$\;\pm\;$0.09 &   35 &   57 & 0.609$\;\pm\;$0.131 \\
n6652                  & -0.67 & -0.75 & 16.50$\;\pm\;$0.02 &
 0.34$\;\pm\;$0.08 &   55 &   97 & 0.567$\;\pm\;$0.096 \\
n6681  (M~70)          & -1.06 & -1.30 & 15.62$\;\pm\;$0.03 &
-0.14$\;\pm\;$0.08 &   46 &   83 & 0.554$\;\pm\;$0.102 \\
n6717  (Pal~9)         & -0.89 & -1.11 & 15.77$\;\pm\;$0.05 &
-0.13$\;\pm\;$0.09 &   16 &   22 & 0.727$\;\pm\;$0.239 \\
n6723                  & -0.79 & -0.88 & 15.65$\;\pm\;$0.03 &
 0.09$\;\pm\;$0.10 &   42 &  136 & 0.309$\;\pm\;$0.055 \\
n6760$^\ast$           & -0.52 & -0.38 & 18.40$\;\pm\;$0.03 &
 0.38$\;\pm\;$0.08 &  158 &  279 & 0.566$\;\pm\;$0.056 \\
n6838  (M~71)          & -0.56 & -0.44 & 14.92$\;\pm\;$0.04 &
 0.37$\;\pm\;$0.08 &   13 &   27 & 0.482$\;\pm\;$0.163 \\
n6864$^\ast$  (M~75)   & -0.89 & -1.11 & 17.76$\;\pm\;$0.03 &
 0.06$\;\pm\;$0.08 &  230 &  415 & 0.554$\;\pm\;$0.046 \\
n6934                  & -1.09 & -1.33 & 16.72$\;\pm\;$0.07 &
-0.24$\;\pm\;$0.11 &   54 &  121 & 0.446$\;\pm\;$0.073 \\
n6981  (M~72)          & -1.09 & -1.33 & 16.86$\;\pm\;$0.04 &
-0.04$\;\pm\;$0.09 &   42 &   67 & 0.627$\;\pm\;$0.123 \\
n7078$^\ast$  (M~15)   & -1.91 & -1.94 & 15.44$\;\pm\;$0.07 &
-0.28$\;\pm\;$0.12 &  129 &  263 & 0.491$\;\pm\;$0.053 \\
n7089  (M~2)           & -1.18 & -1.41 & 15.79$\;\pm\;$0.04 &
-0.24$\;\pm\;$0.09 &   80 &  163 & 0.491$\;\pm\;$0.067 \\
\hline
\hline
\end{tabular}
\end{center}
\end{table*}

\begin{figure*}
\begin{center}
\includegraphics[width=5.9cm]{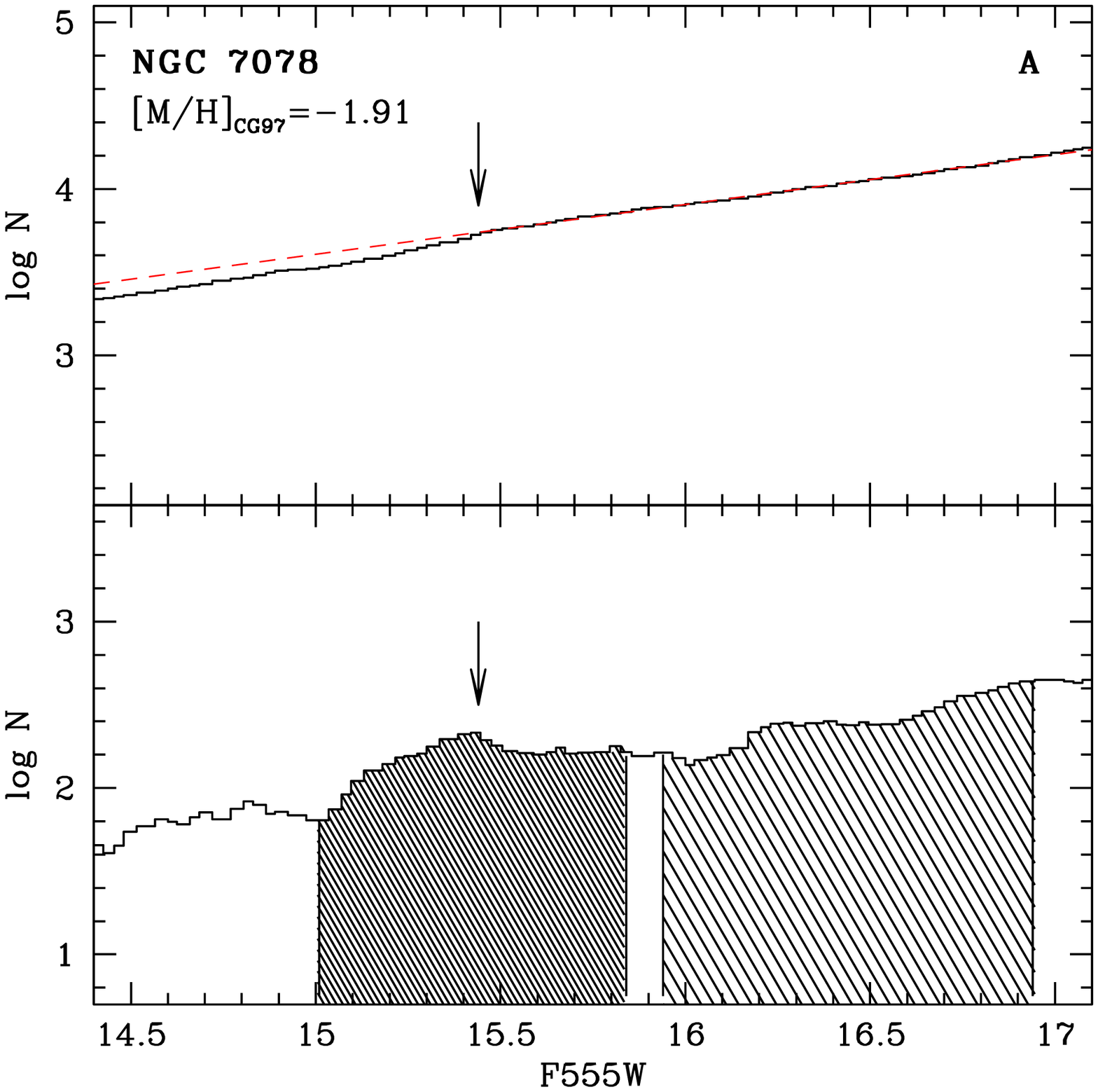}
\includegraphics[width=5.9cm]{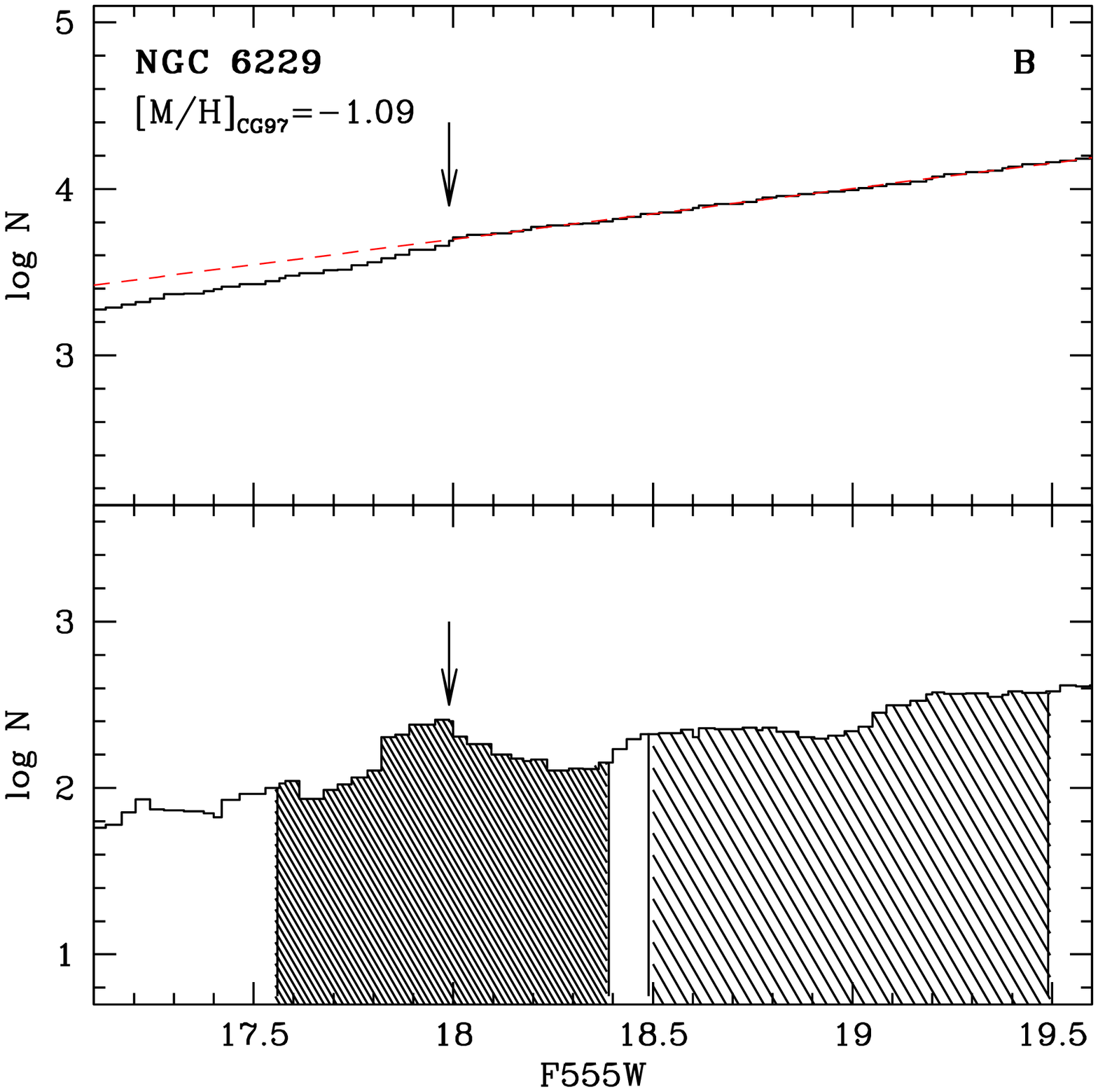}
\includegraphics[width=5.9cm]{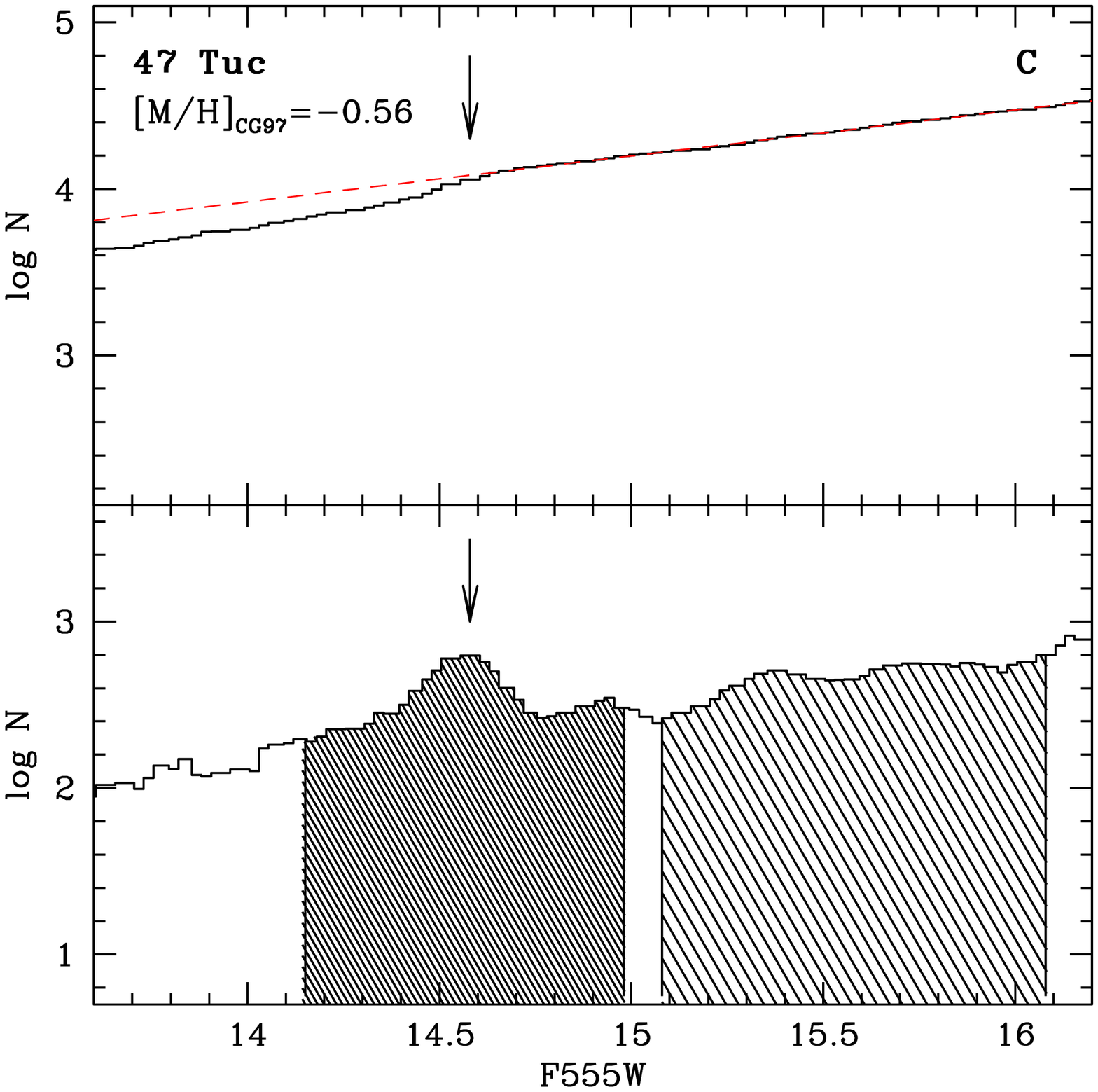}
\caption{Example of the bump detection in three clusters.
{\em Panel A} - The two panels show the cumulative (top) and
differential (bottom) RGB LF for the metal poor cluster NGC 7078;
{\em Panel B} - As in {\it Panel A}, but for the intermediate metallicity
cluster NGC~6229;
{\em Panel C} - As in {\it Panel A}, but for the metal rich cluster 47 Tuc.
The global metallicity on the CG97 scale (see the text for details) is
shown in each panel.
The bump position (marked by the arrow) clearly shows up as a change in the 
slope of the cumulative LF, and it is confirmed by the differential LF.}
\label{LFfig}
\end{center}
\end{figure*}

\subsection{Measurement of the \dfbump\ parameter}

The parameter \dfbump\ used in this work is defined as the difference
between the RGB bump and Zero Age HB (ZAHB) HST F555W magnitudes; it is 
therefore necessary to detect the bump along the cluster RGB and evaluate 
the ZAHB brightness at the level of the RR Lyrae instability strip, 
following Z99.  

To measure the position of the RGB bump we used both the cumulative and 
differential LFs corrected using the completeness functions for the RGB 
sequences obtained from the artificial star experiments (Piotto et al. 
\cite{snapshot}).
For those clusters from which more populous samples of RGB stars were 
obtained (e.g. the 26 clusters marked with an asterisk in Tab.~1) the 
cumulative LF alone provides a solid determination of the bump level, which 
is always consistent with the existence of a statistically significant bump 
in the differential LF (see Fig.~1 for three examples).
In case of the less populated clusters, both cumulative and differential LF 
have to be used simultaneously in order to estimate the bump level, as 
first discussed by Fusi Pecci et al.~(1990). As discussed in sections 4.1 
and 4.2, the bump detections in the more poorly sampled clusters are genuine,
and in general, our detections for the whole sample are internally consistent.

We detected the RGB bump in 54 clusters, whereas for 20 other clusters this 
was not possible due either to poor RGB statistics, or to a large 
differential reddening, or to high field star contamination, especially for 
bulge clusters.

Fig.~\ref{LFfig} shows some examples of the bump identification in the 
differential and in the cumulative LFs for three clusters in our database: 
a metal-poor cluster (NGC~7078, [M/H]$_{\mathrm{CG97}}=-1.91$), an 
intermediate-metallicity (NGC~6229, [M/H]$_{\mathrm{CG97}}=-1.09$), and a 
metal-rich one (47~Tuc, [M/H]$_{\mathrm{CG97}}=-0.56$). In all panels, the 
position of the bump is marked by a vertical arrow.

The error bars on the bump position have been estimated by adding in
quadrature the photometric error at the bump magnitude and the half width 
of the bin size used in the RGB LF.

Estimation of the ZAHB luminosity is a risky procedure due to the variety 
of HB morphologies and also to the fact that we cannot directly evaluate 
the mean magnitude of the RR Lyrae stars from our photometry.
In fact, our photometric data cover only a very short time interval and this 
does not allow a suitable sampling of the pulsation cycle of the variables. 
Therefore, we followed the same procedure adopted by Z99. It can be briefly 
summarized as follows:
\begin{itemize}
\item
clusters with $\feh\le-1$ (metal-intermediate and metal-poor): we divided 
the sample into metallicity groups (each group spans at most 0.2 dex in 
metallicity); to each metallicity group we associated a template cluster 
(with a metallicity within the group); the template clusters have accurate 
photometry and, most importantly, a sizable number of RR Lyrae, so that we 
could obtain an accurate estimate of the RR Lyrae magnitude level. Then we 
registered the HB of each cluster in the group to the HB of the reference 
cluster by artificially shifting the CMD of the template cluster in both 
color and magnitude in order to overlap both their RGBs and HBs.
The mean magnitude of the RR Lyrae for each cluster was then obtained by 
shifting the RR Lyrae magnitude of the reference cluster by the magnitude 
difference adopted to overlap the two CMDs. Finally, we transformed the RR 
Lyrae mean magnitude into the ZAHB magnitude following Cassisi \& Salaris
(1997; see Z99 and Recio-Blanco et al. \cite{zahb} for further details);

\item
clusters with $\feh>-1$ (metal-rich) : we first estimated the photometric 
error ($\sigma_{\mathrm{F555W}}$) at the level of the HB using our set of 
artificial star experiments; then we determined the lower envelope of the 
HB stellar distribution. The ZAHB magnitude was eventually fixed at 
$3\sigma_{\mathrm{F555W}}$ magnitudes above the lower envelope.
\end{itemize}

The individual ZAHB luminosities as well as a more detailed description of 
the adopted procedure will be discussed in a companion paper (Recio-Blanco 
et al. 2003, in preparation).

The errors in \dfbump\  have been estimated by summing in quadrature the 
errors in the bump position $\mathrm{F555W}_{\mathrm{bump}}$ and in the 
ZAHB level $\mathrm{F555W}_{\mathrm{ZAHB}}$.

Because of the uncertainty that still affects the metallicity scale for
GGCs (Rutledge, Hesser \& Stetson~\cite{rhs97}; VandenBerg~\cite{v00};
Caputo \& Cassisi \cite{cc02} and Kraft \& Ivans \cite{ki03}), 
we have used both the Carretta \& Gratton (\cite{cg97}, hereinafter CG97) 
and the Zinn \& West (\cite{zw84}, hereinafter ZW84) metallicity scale.
Moreover, we adopt, due to the lack of individual spectroscopic measurements 
for several GGCs in our sample, a mean $\alpha$-enhancement of 0.3 dex for 
metal-poor and metal-intermediate clusters $(\feh<-1.0)$ and of 0.2 dex for
metal-rich clusters $(\feh>-1.0)$. The former value was suggested by Carney 
(\cite{carney}, C96), while the latter is a mean between the estimates 
collected by C96 and those collected by Salaris \& Cassisi (\cite{sc96}).
For each metallicity scale, the global cluster metallicity was estimated by 
using the relation provided by Salaris, Chieffi \& Straniero~(\cite{scs93}).
We adopt a global metallicity error of $\pm$ 0.15 dex. This value can be
considered as a safe lower limit to the uncertainties affecting both $\feh$ 
and $[\alpha/\mathrm{Fe}]$ measurements (Rutledge et al.~\cite{rhs97}).

Table~\ref{tabump} lists the relevant data for all clusters in our sample 
where the RGB bump has been detected. Column (1) gives the identification, 
columns (2) and (3) give the cluster global metallicity according to the 
metallicity scale provided by CG97, and by ZW84, respectively.
Column (4) lists the F555W magnitude of the RGB LF bump, column (5) lists 
the \dfbump\ values with associated errors.

\subsection{Measurement of the \rbump\ paramenter}

We measured the \rbump\ parameter (using the same definition of B01 but 
the F555W magnitude) for all the 54 GGCs where we detected the bump.
The star counts in the bump and in the normalization regions were corrected 
using the completeness functions given by the artificial star experiments.

The stellar counts in Bump and Normalization area are listed in columns (6) 
and (7) of Tab.~1, respectively. The values of the \rbump\ parameter with 
the corresponding errors are listed in column (8).
The 1$\sigma$ error on the observed \rbump\ values is determined according
to $\sigma$(\rbump)=\rbump\ $\sqrt{(1/N_b)+(1/N_n)}$ where $N_b$ and $N_n$ 
are the star counts in the bump and in the normalization region, respectively.

\section{The theoretical framework}

The theoretical framework we adopt in this investigation is based on an
updated and larger set of stellar models (Pietrinferni, Cassisi, Salaris \& 
Castelli 2003). The new models have been computed by using a recent version 
of the FRANEC evolutionary code. The input physics has been updated with 
respect to the models used in Z99 (Cassisi \& Salaris~1997), and the changes 
are summarized in the following:

\begin{itemize}
\item
the radiative opacity is obtained from the OPAL tables (Iglesias \& Rogers 
\cite{iglesias}) for temperatures larger than $10^4$ K, and from Alexander 
\& Ferguson (\cite{alexander}) for lower temperatures. Opacity for electron 
degenerate matter is computed following Potekhin (\cite{pot}).

\item
We updated the energy loss rates for plasma-neutrino processes by using the 
most recent and accurate results provided by Haft, Raffelt \& 
Weiss~(\cite{haft}). For all other processes we still rely on the same 
prescriptions adopted by Cassisi \& Salaris (1997).

\item
The nuclear reaction rates have been updated by using the NACRE database 
(Angulo et al.~1999), with the exception of the 
$^{12}$C$(\alpha,\gamma)^{16}$O reaction. For this reaction we employ the 
more accurate recent determination by Kunz et al.~(\cite{kunz}).

\item
The detailed Equation of State (EOS) by A. Irwin\footnote{The EOS code is 
made publicly available at ftp://astroftp.phys.uvic.ca under the GNU 
General Public License (GPL)} has been used.
A full description of this EOS is still in preparation (Irwin et al.~2003) 
but a brief discussion of its main characteristics can be found in Cassisi, 
Salaris \& Irwin (\cite{csi03}). It is enough to mention here that this EOS, 
whose accuracy and reliability is similar to the OPAL EOS developed at the 
Livermore Laboratories (Rogers, Swenson \& Iglesias \cite{rsi96}) and 
recently updated in the treatment of some physical inputs (Rogers \& Nayfonov 
\cite{rn02}), allows us to compute self-consistent stellar models in all 
evolutionary phases relevant to this investigation.

\item
The extension of the convective zones is fixed by means of the classical 
Schwarzschild criterion. Induced overshooting and semiconvection during the 
He-central burning phase are accounted for following Castellani et 
al.~(\cite{cast85}). The thermal gradient in the superadiabatic regions is 
determined according to the mixing length theory, whose free parameter has 
been calibrated by computing a solar standard model.

\item
Our evolutionary code includes the process of atomic diffusion of both Helium
and heavy elements; the models used in this work, however, have been computed 
with the atomic diffusion switched off, taking into account that Cassisi et 
al.~(\cite{cds97}) have clearly shown how the effect of this physical process 
on \dvbump is negligible.

\item
We adopt a scaled-solar heavy element mixture (Grevesse \& Noels~\cite{gn93}); 
the enhancement of $\alpha$-elements observed in GGCs is accounted for 
following the prescriptions by Salaris et al.~(\cite{scs93}).

\item
As far as the initial He-abundance is concerned, we adopt the estimate
recently provided by Cassisi et al. \cite{csi03} on the basis of new
measurements of the $R$ parameter in a large sample of GGCs. 
They found an initial He-abundance for GGC stars of the order of $Y=0.245$, 
which is in fair agreement with recent empirical measurements of the 
cosmological baryon density provided by W-MAP (Spergel et al.~2003). To 
reproduce the calibrated initial solar He-abundance we used $dY/dZ\approx1.4$
(Pietrinferni et al.~2003).

\item
Bolometric magnitudes have been transformed into HST F555W magnitudes
according to the transformations provided by Origlia \& Leitherer
(\cite{origlia}), based on the atmosphere models computed by Bessell, 
Castelli \& Plez (\cite{bcp98}).
\end{itemize}
A more detailed discussion about the model input physics and the 
evolutionary code is presented in Pietrinferni et al.~(2003).

Finally, we discuss in more detail the metal distribution used to compute
the evolutionary models. In our analysis we will compare the models computed 
with a scaled-solar metal abundance to GGCs data for which the metal 
abundance is $\alpha$-enhanced. According to Salaris et al.~(1993), as long 
as the $\alpha$-elements have approximately the same enhancement, scaled-solar 
models mimic $\alpha$-enhanced models computed with the same global 
metallicity [M/H].
This is, however, strictly true only for [M/H] values up to $\sim -$1.0, as
shown by Salaris \& Weiss~(1998) and Vandenberg et al.~(2000). For higher 
global metallicities this equivalence is not well satisfied anymore. 
Therefore we investigated how different would be the \dfbump\ parameter at 
a given [M/H] larger than $-$1, for a scaled-solar and an $\alpha$-enhanced 
metal mixture. We used isochrones by Salaris \& Weiss~(1998) with [M/H]=$-$0.3 
(approximately the upper end of the metallicity range spanned by our GGC 
sample), both scaled-solar and with [$\alpha$/Fe]=0.4; we found that the 
\dfbump\ values are changed by only 0.05 mag for typical GGC ages, the 
$\alpha$-enhanced ones being larger. This is however just an upper limit to 
the real difference, since these metal rich clusters seem to show an 
$\alpha$-enhancement lower than [$\alpha$/Fe]=0.4. We conclude that, as far 
as the \dfbump\ parameter is concerned, we can safely use scaled-solar 
models with the cluster global [M/H] even in the high metallicity regime. 
We also verified that the \rbump\ values are unaffected by the 
$\alpha$-enhanced metal distribution.

\begin{table}[!t]
\begin{center}
\caption{Theoretical \dfbump\ values as a function of $[M/H]$ and age.}
\label{tabumpth}
\begin{tabular}{ccccc}
\hline
\hline
\noalign{\smallskip}
$[M/H]$ & \multicolumn{4}{c}{\dfbump} \\
\noalign{\smallskip}
& $10$ Gyr & $12$ Gyr & $14$ Gyr & $16$ Gyr. \\
\noalign{\smallskip}
\hline
-2.267 & -0.907 & -0.846 & -0.770 & -0.699 \\
-1.790 & -0.690 & -0.621 & -0.555 & -0.494 \\
-1.266 & -0.297 & -0.221 & -0.159 & -0.116 \\
-0.963 & -0.056 &  0.004 &  0.056 &  0.108 \\
-0.659 &  0.192 &  0.271 &  0.347 &  0.418 \\
-0.253 &  0.469 &  0.566 &  0.649 &  0.694 \\
 0.000 &  0.639 &  0.741 &  0.791 &  0.843 \\
\hline
\hline
\end{tabular}
\end{center}
\end{table}

\begin{figure*}[!htp]
\begin{center}
\includegraphics[width=8.5cm]{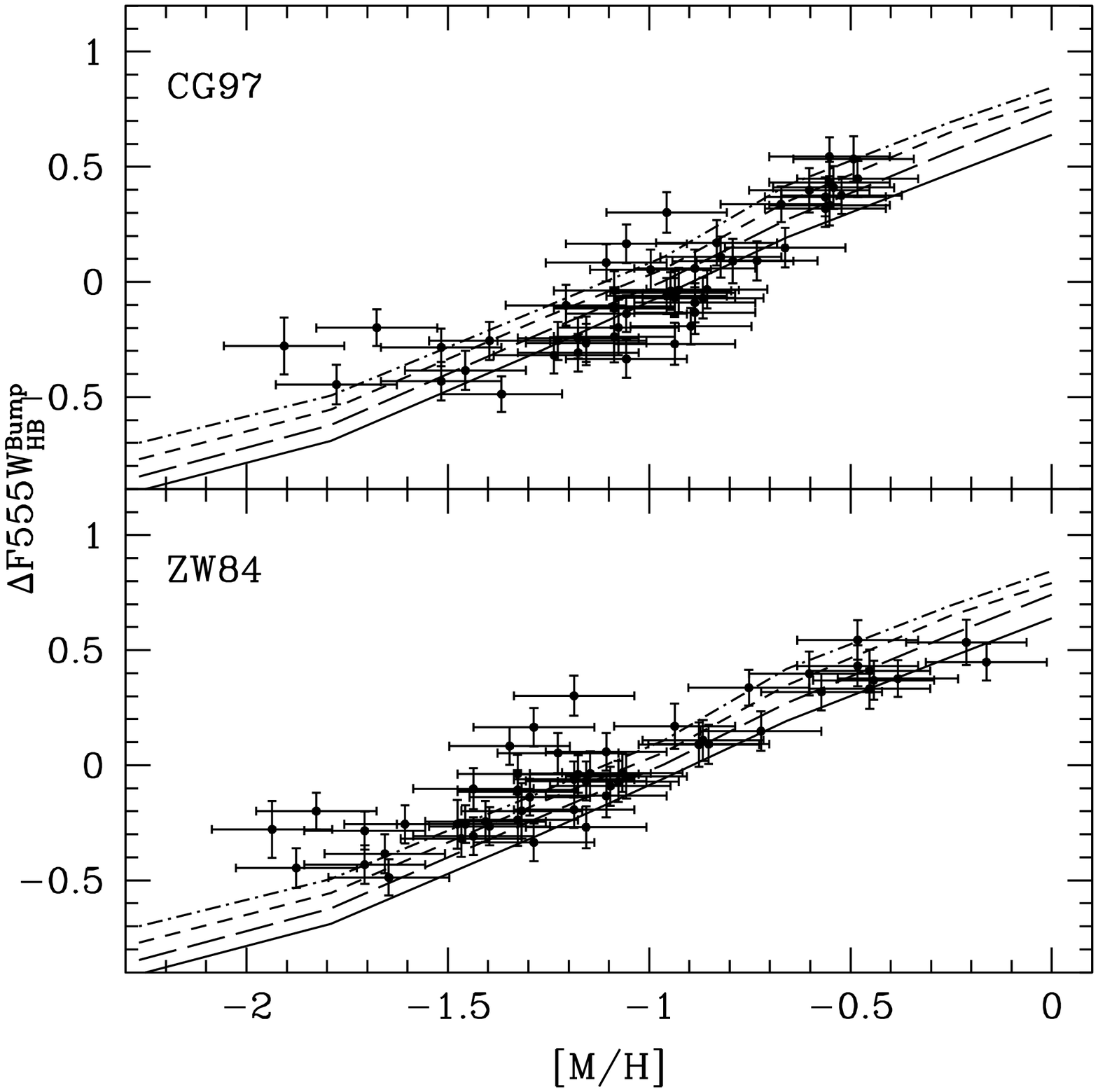}
\includegraphics[width=8.5cm]{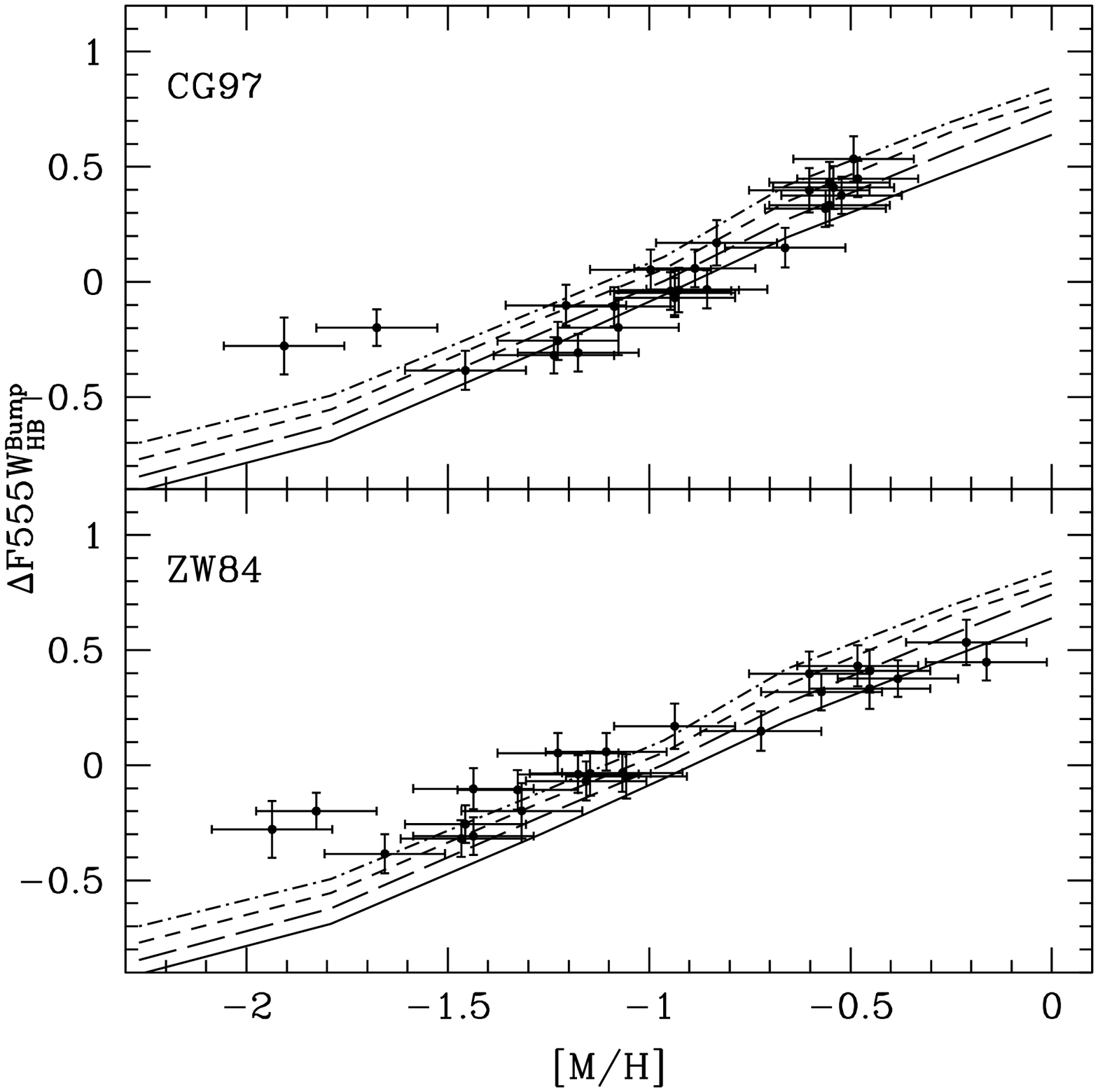}
\caption{Comparison between the theoretical and empirical values of \dfbump\ 
as a function of the global metallicity for all the 54 GGCs in our sample 
(left panel) and for the 26 GGCs with more than 85 stars in the bump region 
(right panel). The empirical \dfbump\ data are plotted as a function of both 
the Zinn \& West (\cite{zw84} -- ZW84) and Carretta \& Gratton (\cite{cg97} 
-- CG97) metallicity scales (see text for more details). The theoretical
predictions are plotted for four different cluster ages, namely 10 Gyr 
(solid line), 12 Gyr (long dash), 14 Gyr (short dash), and 16 Gyr (dot - 
short dash).}
\label{modelbump}
\end{center}
\end{figure*}

\section{Comparison between theory and observations}
\label{th2obs}

\subsection{The \dfbump\ parameter}

Figure~\ref{modelbump} shows the comparison between theoretical predictions 
(given in Tab.~2) and measured values of \dfbump\, as a function of the 
global metallicity [M/H], for a selected age range.  
The behaviour of the theoretical \dfbump\ stems from the strong dependence 
of the depth of the convective envelope on both metallicity and mass. The
higher the cluster age, the smaller the evolving mass along the RGB, with an 
ensuing deeper extension of the convective envelope, and a fainter bump 
brightness, whereas the HB level is largely unaffected.
On the other hand, at a fixed age, the larger the metallicity, the deeper 
the extension of the convective envelope which causes a fainter bump; the 
HB is also fainter for increasing metallicity, but the effect on the bump 
level is larger, thus causing an increase of \dfbump.

The left panel of Fig.~2 shows the data for the entire sample of GCs, 
whereas in the right panel we plot only the clusters with more than 85 
stars in the bump region (hereafter we will refer to this sample of 
clusters as the best cluster sample). A glance at this figure reveals that 
the distribution of the observational data for the entire sample closely 
resembles the distribution for the best cluster sample which, incidentally, 
spans the entire metallicity range covered by the full sample. This evidence 
supports the significance of the bump detections for the whole 54 clusters 
listed in Tab.~1.  
Further confirmation comes from the fact that, as we will see in the
following, the comparison with theory provides exactly the same results when 
considering either the full sample or the best cluster sample.
We anticipate that the analysis of the \rbump\ parameter, presented in the 
next subsection, will provide additional evidence for the significance of 
the bump detection in all the clusters displayed in Table~1.

Figure~\ref{modelbump} shows also a qualitative agreement between 
observations and theory. In particular, both theoretical lines and 
observations lie in the same region of the \dfbump\-[M/H] plane, and the 
observed trend with metallicity is well reproduced by theory (with the 
possible exception of the three most metal poor clusters).

We have also performed a more quantitative comparison, though it must be 
noted that it is made difficult (and uncertain) by the strong dependence 
of \dfbump\ on the cluster age and metal content.
Indeed, \dfbump\ varies by 0.03 magnitudes for a 1 Gyr variation in age,
and by the same amount for a variation of only 0.04 dex in [M/H].
To assess the internal accuracy of theoretical models we compared the 
cluster age required to fit the observed \dfbump\ values with the age 
determinations obtained from the main sequence (MS) turn off (TO) position.
It must be explicitly noted that we do not propose to use the \dfbump\
parameter as an age indicator; however, we should expect that, for any given 
metallicity scale, the age scale inferred from the \dfbump\ agrees with the 
ages from the TO.
Systematic or random age offsets imply (systematic or random) effects not
properly accounted for by the models.

\begin{figure}[!t]
\begin{center}
\includegraphics[width=9cm]{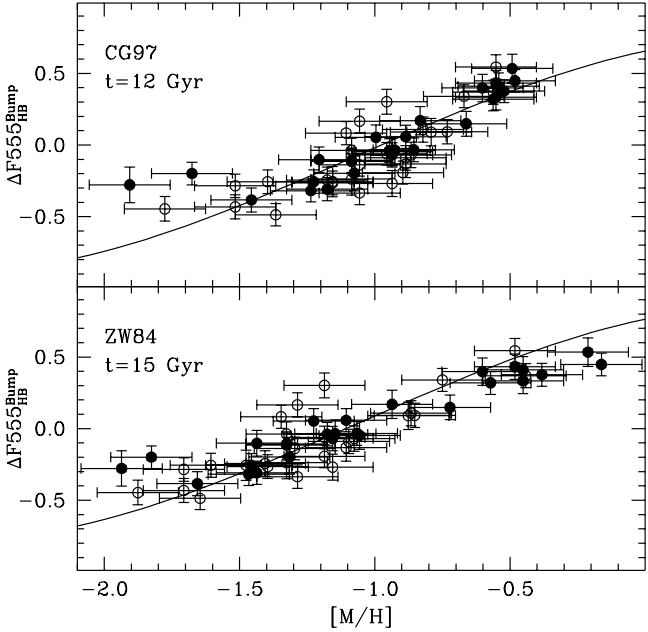}
\caption{Comparison between observed and predicted \dfbump\ values for the 
labeled average ages determined in case of, respectively, the CG97 (upper 
panel) and ZW84 (lower panel) [M/H] scale. Filled circles denote the 26 
GGCs with more than 85 stars in the bump region, whereas open circles 
represent the remaining clusters in our sample.}
\label{DFconf}
\end{center}
\end{figure}

We started with the CG97 metallicity scale. The theoretical line which best 
fits the observed distribution of \dfbump\ with the metallicity has an age 
of $11.8$ Gyr, with a 1$\sigma$ dispersion of 4.0 Gyr.  Figure~\ref{DFconf}
(upper panel) displays the comparison between the observed \dfbump\ and the 
models for the labeled age. The use of only the best cluster sample (26 
objects) sample does not affect the age, though the spread is slightly 
reduced. A close inspection of Fig.~\ref{DFconf} shows that clusters with 
[M/H]$< -$1.6 are systematically discrepant, even those with the best 
populated bump. This fact may point out to some serious inaccuracy of the 
stellar models for this metallicity range, though we cannot exclude that the 
discrepancy may be ascribed to a different \lq{true}\rq\ global metallicity 
for these clusters (e.g., the $\alpha$-enhancement is higher than assumed). 
Moreover, the discrepant clusters in this metallicity range are only a few, 
namely M~15, M~68, and M~53. A larger sample of metal-poor clusters with a
well-populated RGB is required before a definitive conclusion on this
possible discrepancy can be reached.

If we restrict our analysis to the clusters with [M/H]$> -$1.6, the best 
fitting model corresponds to an age of $\approx 11$ Gyr, which does not 
change when we consider only objects belonging to the best cluster sample.
The dispersion of the data points around this best fitting model does,
however, decreases from 3.4 to 2.6 Gyr when using the best sample. Most 
interestingly, we also found that the age scatter around the best fitting
model does not correlate significantly with [M/H].

A different result is obtained using the ZW84 metallicity scale. In this 
case, the age of the best fitting model is $\approx 15$ Gyr 
(Fig.~\ref{DFconf}), but now the residuals show a correlation with [M/H].  
In order to obtain a better match with the theoretical \dfbump\, a younger 
age for more metal-rich clusters must be used.
More specifically, we find an overall statistically significant age-[M/H] 
relationship, with a slope $(-5.7\pm1.1) \rm Gyr~dex^{-1}$.
If we neglect the three most metal-poor clusters mentioned above (they have 
[M/H]$< -$1.8 in the ZW84 scale), we obtain a slope
$(-4.8\pm1.1) \rm Gyr~dex^{-1}$, which does not change for the best cluster 
sample (with [M/H]$> -$1.8). In this case, the best fitting model 
correspond has an age of $\approx 14$ Gyr.

In summary, when using the CG97 scale the cluster average age is of 11-12 
Gyr, in agreement with the TO ages obtained by Salaris \& Weiss (2002), who
determine the GGC distances from their theoretical models.
Using the same metallicity scale, and measuring distances with the classical 
MS fitting technique, Carretta et al.~(2000) find a similar age (though 
their best age estimate is roughly 13 Gyr, when they consider also other 
independent distance scales)
\footnote{Although a detailed discussion of the GGCs ages is clearly beyond 
the scope of this investigation, it has to be mentioned that a very recent
estimate (Gratton et al. \cite{g03}) of the absolute age of one
metal-intermediate and one metal-poor cluster, based on local subdwarfs,
and on new, accurate determinations of reddening and metallicity, is
between 13.5 and 14 Gyr.}.
A second result of the comparison shown in Fig. 3 is that, when using the 
CG97 metallicity scale, we do not find any statistically significant
age-metallicity trend from the \dfbump\ parameter, whereas Rosenberg et 
al.~(\cite{alfred}), VandenBerg~(2000), Salaris \& Weiss~(2002), find in 
general that the more metal rich clusters are younger, and Gratton et 
al.~(\cite{g03}) find that the metal-rich cluster 47~Tuc is about 2.6 Gyr 
younger than more metal-poor clusters like NGC~6752.

Salaris \& Weiss~(2002) obtain, on average, ages higher by about 1 Gyr when 
moving from the CG97 to the ZW84 metallicity scale, but still lower than the 
age derived from the \dfbump. VandenBerg~(2000) determines ages of about 14 
Gyr for the oldest metal-poor GGCs using his own HB models and the ZW84 scale,
and ages decreasing on average when moving to the higher metallicity regime. 
These results confirms the earlier work by Rosenberg et al.~(\cite{alfred}),  
who find that the most metal rich clusters are 15-20\% younger than the most 
metal poor ones. In conclusion, the age-metallicity relationship we found 
from \dfbump\ when adopting the ZW84 metallicity scale, is in qualitative 
agreement with the results we obtain from the TO ages, but the slope is 
larger. Also the average age implied by the \dfbump\ parameter seems larger 
than the most recent determinations.

Unfortunately, this analysis provides somewhat not definitive results about 
the accuracy of the theoretical \dfbump\ values. This is due mainly to the 
current uncertainties in the cluster metallicities and, to a minor extent, 
to uncertainties in the GGC ages (which depend themselves on the metallicity
errors). In fact, typical differences among the TO absolute GGC ages 
mentioned before are of the order of 1-2 Gyr; these differences affect 
\dfbump\ (at a given [M/H]) at the level of 0.03-0.06 mag, whereas typical 
differences of 0.2 dex between the CG97 and ZW84 [M/H] scale modify \dfbump\ 
(at a fixed age) by $\sim$0.2 mag.

\begin{figure}
\begin{center}
\includegraphics[width=9cm]{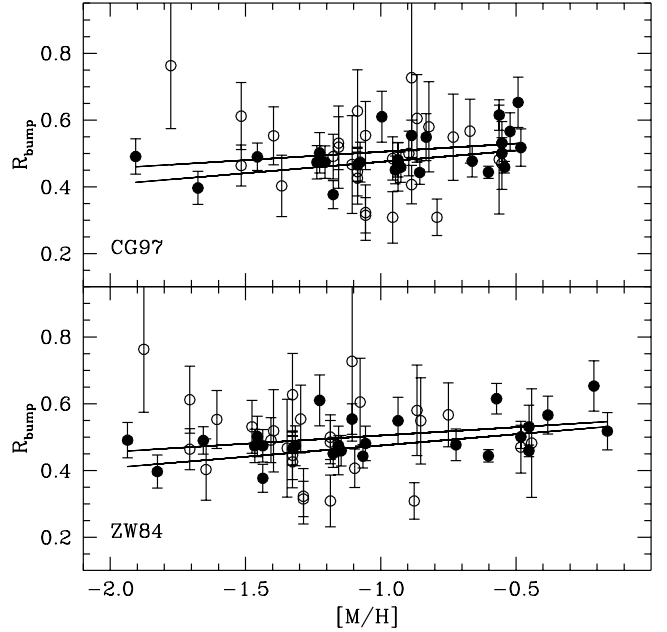}
\caption{Comparison between the theoretical and empirical values of \rbump\ as 
a function of the global metallicity for all the GGCs in our sample; symbols 
are as in Fig.~\ref{DFconf}. The bottom and upper panels refer, respectively, 
to the ZW84 and CG97 [M/H] scale. The solid lines show the theoretical values 
for 10 (upper line) and 16 Gyr.}
\label{modelrbump}
\end{center}
\end{figure}

\subsection{The \rbump\ parameter}

The comparison between theoretical predictions and empirical measurements
of the \rbump\ parameter is shown in Fig.~\ref{modelrbump}, for the same 
two metallicity scales discussed before.
It is important to remark that, as in case of \dfbump\, the \rbump\ 
distribution for the clusters with the most populated bump region does not 
show any evident systematic offset when compared to the full sample. Due to 
the fact that one expects a weak dependence of \rbump\ on both [M/H] and 
age, it is meaningful to compare the mean measured \rbump \ values for the 
best cluster sample and the full sample.
The mean value of \rbump\ for the latter is 0.498$\pm$0.092 (1$\sigma$ 
dispersion); in case of selecting only the 26 objects belonging to the best 
cluster sample (which corresponds to clusters with $\sigma$(\rbump)/\rbump
below 13\%) we obtain 0.497$\pm$0.064, practically identical to the value 
for the whole sample, but with a smaller dispersion because of smaller 
individual error bars (due to Poisson fluctuations in the values of $N_b$ 
and $N_n$).
This is another clear evidence for the significance of our bump detections. 
In fact, in case of spurious detections in the less populated clusters, 
caused by random fluctuations in the number of RGB stars, we would expect to 
obtain an average \rbump\ value equal to $\sim$0.37 (corresponding to models 
where the H-abundance jump is absent) with a given dispersion around this 
value due to Poisson noise; however, when we consider only the less populated
clusters with $\sigma$(\rbump)/\rbump\ larger than 13\%, we obtain a 
distribution of points with a mean value of 0.498 equal to the best cluster 
sample, and not centred around 0.37.

To test the consistency of the observed \rbump\ with theory, we adopted the 
following approach: for a given metallicity scale we computed, on a cluster 
by cluster basis, the corresponding theoretical \rbump\ values, using the 
ages obtained from \dfbump. The precise age value is however not crucial, 
due to the very low sensitivity of the \rbump\ parameter to age (see the 
discussion in B01 and data plotted in Fig.~\ref{modelrbump}).
Also the dependence on [M/H] is very weak, as noticed before.

\begin{figure}
\begin{center}
\includegraphics[width=9cm]{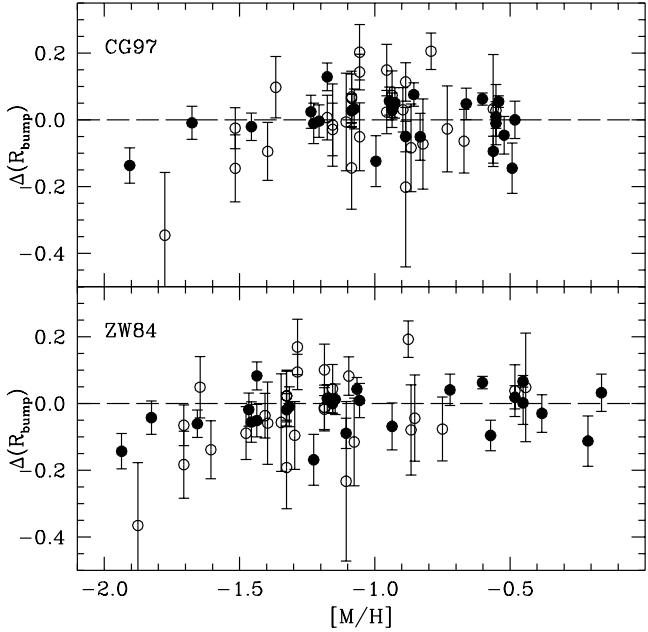}
\caption{Difference between the theoretical \rbump\ values computed for the 
individual cluster ages determined from \dfbump\, and the observational 
counterpart, in case of the CG97 (upper panel) and ZW84 (lower panel) [M/H] 
scale. The horizontal dashed line corresponds to a zero difference and it is 
not a fit to the data.}
\label{Rbconf}
\end{center}
\end{figure}

These 54 theoretical values have been compared with the empirical data and 
the difference between theory and observations has been analyzed (see 
Fig.~\ref{Rbconf}). We found an average difference (theory-observations) 
$\Delta$(\rbump)=$-0.002\pm$0.093 (1$\sigma$) for the whole GGC sample when 
using the CG97 metallicity scale, with no evident correlation of the 
residuals with [M/H]. We also tested, making use of Monte Carlo simulations, 
if the dispersion around the mean can be explained by the Poisson errors on 
the empirical determination of \rbump. More in detail, for each cluster we 
randomly generated 10000 values of $\Delta$(\rbump), centred around $-$0.002 
and with a Gaussian 1$\sigma$ dispersion equal to the observational error 
on \rbump. We then joined together the synthetic $\Delta$(\rbump) values for 
each individual cluster, and determined the 1$\sigma$ dispersion of the 
resulting distribution, which compares well with the observed value of 0.093.
We also considered, as a further test, the subsample of 26 clusters with the 
best populated RGB up to the bump region.
The \rbump\ residuals show again a negligible difference from zero, the 
average value being $-0.002\pm$0.065, and once again we found that the 
dispersion around this value can be explained as observational errors on 
the \rbump\ parameter. Note the much smaller value of the dispersion when 
only the best \rbump\ determinations are accounted for. We obtain the same 
result if M~15 and M~53 -- the two most metal poor clusters -- are 
neglected (see Fig.~\ref{Rbconf}).

When using the ZW84 scale, and the age distribution obtained from \dfbump, 
the average value of the difference between theoretical and observed \rbump\ 
is slightly higher, namely $-0.028\pm$0.093, the residuals showing no 
trend with [M/H]. The average value is $-0.022\pm$0.064 if we consider the 
best cluster sample.

\subsection{The effect of chemical self pollution}

To explain the observed CNO abundance anomalies and the extended blue HB 
tail in some GGCs, D'Antona et al.~(2002)  suggested the existence of an 
He-enriched stellar component within the clusters, due to chemical pollution 
by the ejecta of massive asymptotic giant branch stars. We tested whether 
this He-enhancement might affect predicted \dfbump\ and \rbump\ values when 
compared to the case of a constant He-abundance. Following D'Antona et 
al.~(2002) results, we have produced a synthetic CMD for an hypothetical 12 
Gyr old GGC with [M/H]=$-$1.3. We considered that 64\% of the coeval stellar 
population is composed of stars with the standard He-abundance adopted in 
our models, whereas 36\% of the population is made of stars with Y randomly 
distributed between the standard value and an abundance 0.06 higher.
We found that the location of the bump in the LF of this He-enhanced 
population does not show any significant difference with respect to the 
standard case. The HB level at the RR~Lyrae instability strip -- as shown by 
D'Antona et al.~(2002) -- is determined by the 'He-normal' population, and 
therefore, the \dfbump\ value for a cluster with He-enhanced stars is 
expected to be very close to the case of a standard He-normal population.
It is also important to notice that, if the He-enhanced stars have all the 
same abundance 0.06 higher than the bulk of the cluster population, the 
cluster LF would show a second bump, about 0.13 mag brighter than the main one.

As far as the \rbump\ parameter is concerned, we found only a a marginal
difference. In conclusion, the occurrence of chemical pollution due to the 
same GGC stars does not affect the \dfbump\ and the \rbump\ values predicted 
by canonical stellar models.

\section{Summary and final remarks}

In this investigation we adopted a database of HST photometric data for a 
large sample of GGCs, and we have been able to measure the RGB bump location 
in a sample of 54 clusters. For the same cluster sample we have also 
determined the star counts in the bump region following B01. This represents, 
so far, the largest database of empirical estimates for both the RGB bump
brightness and star counts in this evolutionary phase.

The observed magnitude difference between the bump and the HB, i.e. \dfbump\, 
and the star counts in the bump region, i.e. the \rbump\ parameter, have 
been compared with theoretical predictions by using a new set of stellar 
evolution models.
To account for the current uncertainty in the GGC metallicity scale, we have 
adopted both the CG97 [Fe/H] scale and the ZW84 one. We also employed 
reasonable assumptions for the $\alpha$-element enhancement in GGCs stars, 
based on the presently available estimates.

Owing to the sensitivity of the \dfbump\ parameter on cluster age, the ages 
required to fit the observed \dfbump\ values should agree with recent 
independent estimates based on the luminosity of the TO, for the theoretical 
models being consistent with observations. Our results can be summarized as 
follows:
\begin{itemize}

\item
The mean age needed to obtain agreement between observed and predicted 
\dfbump\  values is $\sim 12$ Gyrs (with a $1\sigma$ dispersion of 4.0 
Gyrs) when the CG97 [Fe/H] scale is used. This value is slightly smaller 
than recent independent estimates obtained by using the same metallicity 
scale. It is also worth mentioning that a significant discrepancy exists 
for the most metal-poor clusters ([M/H]$\le-1.6$) in our sample. Current 
data do not allow us to conclusively assess whether this represents a real 
problem for the theoretical models, or the discrepancy is due is an 
observational bias due to the limited sample of stars located along the RGB.
At variance with results from studies of TO ages, the \dfbump\ of metal 
rich clusters like 47~Tuc is reproduced by models with the same average age 
as more metal poor ones.

\item
The mean cluster age is $\sim 15$ Gyr when the ZW84 [Fe/H] scale is used 
and, in addition, we find a statistically significant correlation between 
the residuals around this age and the cluster [M/H]. This mean age of 15 
Gyr appears larger than similar mean GGC ages available in the literature.
The slope of the age-metallicity relationship, though in qualitative
agreement with current estimates (e.g., Rosenberg et al. \cite{alfred}
and Gratton et al. 2003), is also too steep.

\end{itemize}

On the basis of these results we can draw the following conclusion:  
even though a qualitative agreement between theory and observations of 
\dfbump\ does exist, a more definitive assessment of the confidence level 
is hampered by the not negligible uncertainties still affecting both the 
cluster [Fe/H] and [$\alpha$/Fe] estimates.
More robust conclusions require more accurate spectroscopic measurements of 
these quantities.

As far as the \rbump\ parameter is concerned, our analysis suggests that 
there is a good agreement between theory and observations, regardless of the 
adopted metallicity scale. This result is also due to the weak dependence of 
this parameter on the cluster metallicity (and age), which minimizes the 
effects related to the uncertainties on the [M/H] scale.

Finally, we found that the effect on both \dfbump\ and \rbump\ of a possible 
He-rich component -- as recently suggested by D'Antona et al.~(2002) --  in 
the cluster stellar population is negligible.

\begin{acknowledgements}
This work has been partially supported by the Italian Ministero
dell'Istruzione e della ricerca (PRIN2001 and PRIN2002) and by the Agenzia 
Spaziale Italiana. We warmly thank the referee, P. Bergbusch, for his comments 
which greatly improved the presentation of the paper.
\end{acknowledgements}

\end{document}